\documentclass[12pt]{article}  

\usepackage{amsmath,amsfonts,amssymb}
\usepackage{graphicx}
\usepackage{float}
\usepackage[colorlinks=true, allcolors=blue]{hyperref}
\usepackage{makecell}

\providecommand{\keywords}[1]
{
  \small	
  \textbf{\textit{Keywords---}} #1
}

\title{Design of a custom wideband camera for MISTRAL imager-spectrograph}
\author{Eduard R. Muslimov$^{a,b}$,Jerome Schmitt$^{c}$,\\
Christophe Adami$^{b}$,Michel Dennefeld$^{d}$,\\
Marc Ferrari$^{b,c}$ \\
\small a--Department of Physics, University of Oxford, \\
\small Keble Rd, OX14 3RH Oxford, UK\\
\small b--Aix Marseille Univ, CNRS, CNES, LAM, Marseille, France\\
\small c--Observatoire de Haute-Provence, Universite d’Aix-Marseille, CNRS,\\
\small 04870 St-Michel L’Observatoire, France\\
\small d--CNRS, Inst. d’Astrophysique de Paris (IAP) and\\
\small Sorbonne Université (Paris 6),\\
\small 98bis, Boulevard Arago, 75014 Paris, France\\}

\begin{document} 
\maketitle{}

\begin{abstract}
MISTRAL is a visible and near infrared imager and spectrograph working with the $1.93m$ telescope at L'Observa- toire de Haute-Provence. The goal of the present project is to design and build one custom lens covering the entire working band $370-1000 nm$ with an enhanced throughput and resolution. The proposed design has the focal length of $100 mm$ with $f/\#=2$ and consists of 5 lenses with 2 aspheres. It is capable to work in spectroscopy or direct imaging mode with the spectral resolving power up to $R590-1675$ or energy concentration of $84\%$ within $\pm1 pix$. The throughput varies from $79 to 98\%$ in the main band of 400-1000 nm with a commercial AR coating and could be yet improved with a custom one. We also demonstrate that with this image quality can be maintained in  a $\leq 10\%$ margin with practically reachable tolerances.
\end{abstract}


\section{Introduction}
\label{sec:intro}  

Multi-purpose InSTRument for Astronomy at Low-resolution (MISTRAL) is a visible and near infrared EFOSC type imaging spectrograph developed for
the 1.93m telescope at l'Observatoire de Haute Provence \cite{schmitt2024}. It was developed to cover such science cases as:
\begin{itemize}
    \item Follow-up of large variable sky surveys, both current and coming (e.g. Vera C. Rubin
Telescope\cite{Rubin} and SVOM\cite{SVOM} missions),
\item Gamma-ray bursts follow-up,
\item Classification of supernovae,
\item Study of luminous blue variables,
or the numerous peculiar binaries,
\item Photometric and spectroscopic studies of objects, already represented in catalogues, but missing resolution or spectral coverage (especially in the high-latitude regions), etc. 
\end{itemize} 

MISTRAL has been commissioned in 2021 can be operated in two modes: regular observing
runs in visitor mode, and target of opportunity mode with service observing for fast transients

Historically and mainly for the cost reasons, the instrument uses a number of commercial components. In particular,it uses to interchangeable commercial objective lenses to cover the full spectral working range. Namely, Nikkor 105mm f/1.4\cite{Nikkor} lens is used for the visible (VIS) part and a Schneider 
100mm f/2.9\cite{Schneider} lens for the near infrared (NIR) range of 600-1000 nm. This therefore implies a manual change
 of the objective according to the wavelength range of interest. Another issue is that the Schneider lens has a smaller aperture, which causes both obscuration and field vignetting. Finally, these lenses are not optimized to work at the edges of the instrument spectral range, so they may under-perform in terms of both throughput and aberrations. 
 
 Therefore the goal of current work is to develop a custom camera lens for MISTRAL covering the whole 400-1000 nm band at once in either imaging, VIS spectroscopic or NIR spectroscopic modes. It should have 100 mm focal length and $f/\#=2$ aperture to match the incoming beam. The required imaging quality is defined by the  detector -- the instrument  uses Andor ikon-L CCD camera with  $2048 \times 2048 \times 13.5 \mu m$ pixels. So, the lens resolution should be below this value. The objective lens transmission should be maximized in the working region and, if possible, should allow to extend it in the shortwave direction below 400 nm. Below we consider the general optical design of such a camera, demonstrate its' performance and compare  it with the commercial lenses currently in use and also show the technological feasibility of the design.

\section{Camera optical design}
\label{sec:design}  

\subsection{Nominal design}
\label{sec:nominal}

The camera lens under consideration is installed after the intermediate pupil in the spectrograph coupled with 1.93-m telescope. The general view of the full optical system is shown below in Fig.~\ref{fig:lay}. Note, that there are 2 grisms, which can be mounted in position 7, forming low resolution spectra with reciprocal linear dispersion of $\approx15 nm/mm$ in the sub-ranges of 400-800 and 600-1000 nm, respectively. Alternatively,  an empty diaphragm can be mounted in this position, thus implementing an imaging mode. 

\begin{figure} [!ht]
   \begin{center}
   \begin{tabular}{c} 
   \includegraphics[width=0.9\textwidth]{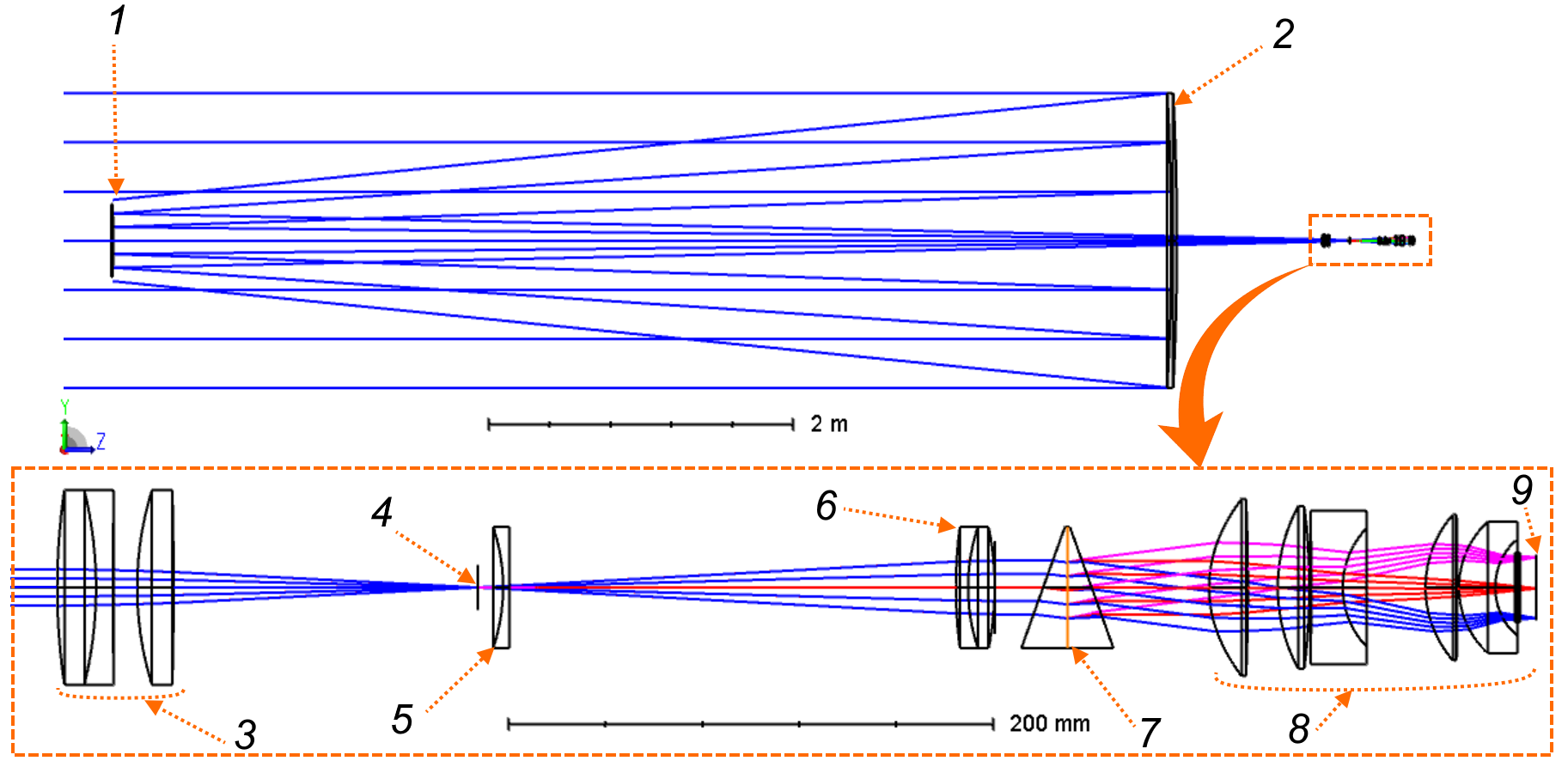}
   \end{tabular}
   \end{center}
   \caption[lay] 
   { \label{fig:lay} 
General view of the instrument optical design: 1  -- telescope primary mirror, 2 -- secondary mirror, 3 -- focal reducer lenses, 4 -- focal plane/ entrance slit, 5 --  field lens, 6 -- spectrograph collimator, 7 -- grism, 8 -- camera lens, 9 -- spectral image plane.}
   \end{figure} 

The custom lens design used a well-known wide-band triplet design as a starting point. Additional lenses were introduced to reach the required $f/\#$. Cemented surfaces are excluded to simplify the manufacturing and assembly. Further, $6^{th}$ order even aspherical surfaces were introduced on the first and the last surface, which are found to be the optimal aspheres positions. Finally, the starting -point design was based on $CaF_2$ as it has low dispersion and high transmission in a wide spectral band. However, this material is expensive and difficult to polish, especially in a case of fast positive lenses, because of its' brittleness. Therefore it was replaced by S-FPL53 glass from OHARA catalogue (see positive lenses L1,2L2 and L4). The remaining 2 meniscus lenses are made of N-KZFS11 and LAK16A from Schott catalogue, respectively.

\begin{figure} [!ht]
   \begin{center}
   \begin{tabular}{c} 
   \includegraphics[width=0.9\textwidth]{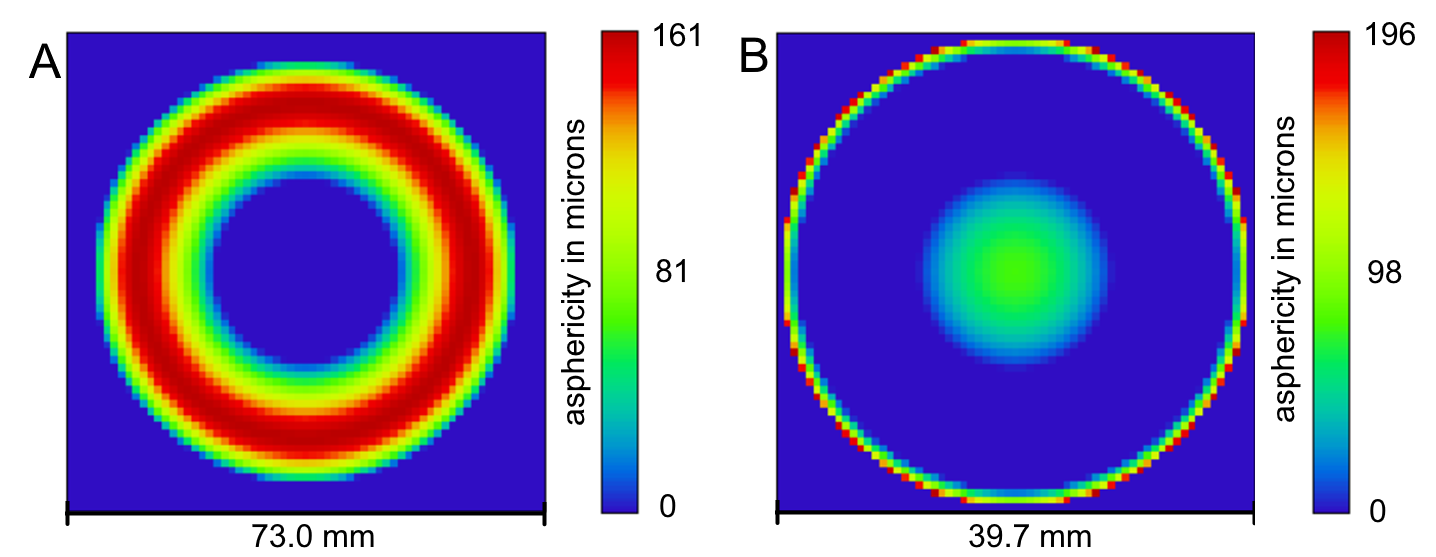}
   \end{tabular}
   \end{center}
   \caption[asph] 
   { \label{fig:asph} 
Surface shape deviation from the best fit sphere: A -- lens 1, surface 1, B -- lens 5 surface 2.}
   \end{figure}

The aspherical surfaces shapes is shown as a deviation from the best fit sphere (BFS) in Fig.~\ref{fig:asph}. The first surface on panel \textit{"A"} has $R_{BFS}=54.29mm$ and the root-mean square (RMS) deviation of $117.5 \mu m$. The last surface in panel \textit{"B"} has $R_{BFS}=26.73mm$ and the RMS deviation $69.4 \mu m$. These values may be challenging, but reachable with the currently available technologies\cite{Beutler16}.

\subsection{Optical performance}
\label{sec:IQ}

\begin{figure} [!ht]
   \begin{center}
   \begin{tabular}{c} 
   \includegraphics[width=0.65\textwidth]{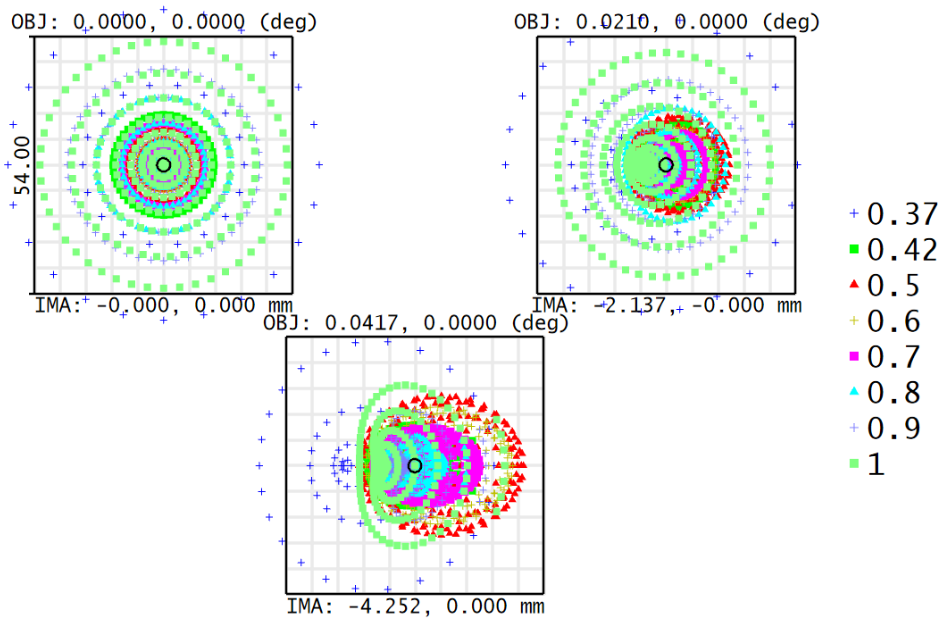}
   \end{tabular}
   \end{center}
   \caption[spot] 
   { \label{fig:spot} 
Polychromatic spot diagrams in the imaging mode at 4 pixels scale.}
   \end{figure}

The geometrical spot diagrams RMS radii were used as the optimization target as the simplest option. So, the first image quality estimation is performed with the spot diagrams. The diagrams in the imaging mode for a lens coupled with the telescope optical system and and the collimator are shown in Fig.~\ref{fig:spot}. The box corresponds to a $4 \times 4 $ pixels area. The numerical values characterizing the geometrical aberrations are also shown in Table~\ref{tab:direct} below.

\begin{figure} [!ht]
   \begin{center}
   \begin{tabular}{c} 
   \includegraphics[width=0.8\textwidth]{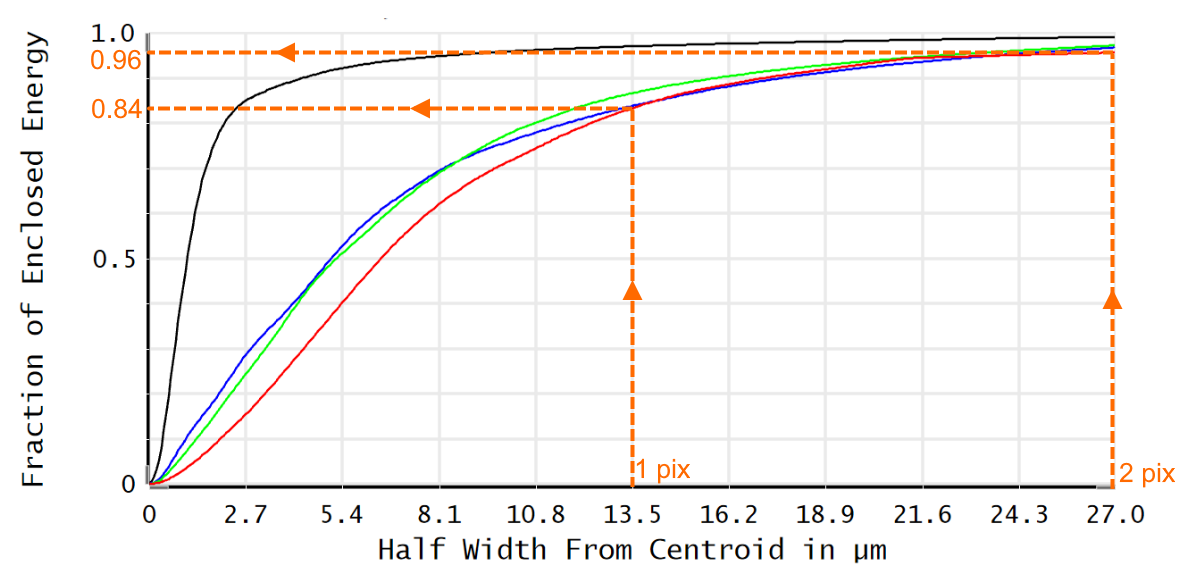}
   \end{tabular}
   \end{center}
   \caption[ener] 
   { \label{fig:ener} 
Ensquared energy in polychromatic PSF in the imaging mode, compared to 1 and 2 pixels radial distance.}
   \end{figure}

   However, the geometrical spots do not provide a correct understanding of the energy distribution in the image. So, we extract the ensquared energy diagrams, corresponding to $2 \times 2$ and $4 \times 4$ are - see Fig.~\ref{fig:ener}. Similarly, the key numerical values are summarized further in Table~\ref{tab:direct}.

\begin{figure} [!ht]
   \begin{center}
   \begin{tabular}{c} 
   \includegraphics[width=0.85\textwidth]{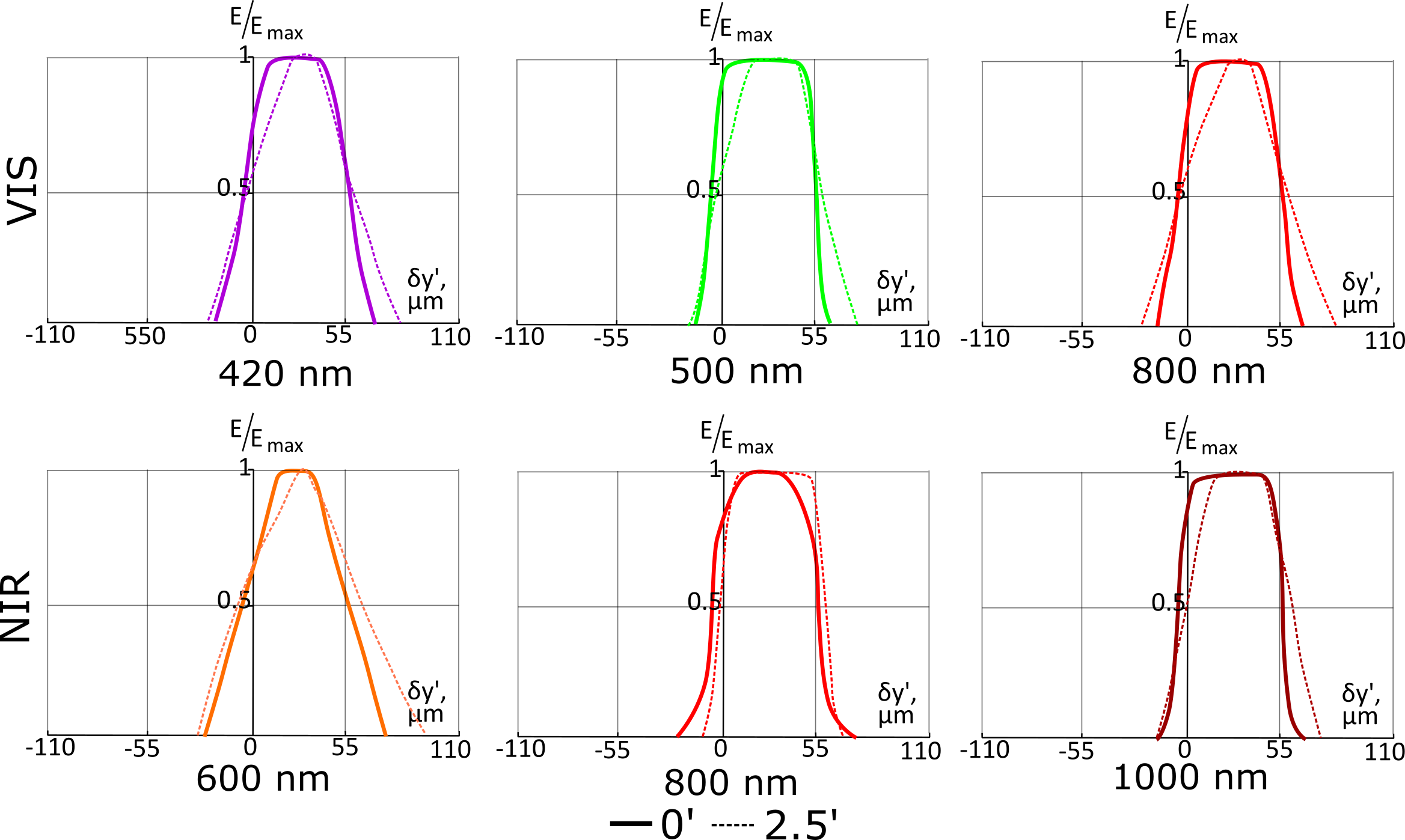}
   \end{tabular}
   \end{center}
   \caption[ifs] 
   { \label{fig:ifs} 
Spectrograph instrument functions in spectral modes with the custom camera lens (geometrical slit image width is $56 \mu m$).}
   \end{figure}

   For the spectral imaging performance we compute the instrument functions, which represent a convolution of the entrance slit brightness distribution and the spectrograph line spread function (LSF) in the dispersion plane. The slit brightness is supposed to be uniform in this case. In practice it may differ from this simple shape significantly depending on the object type, atmospheric turbulence etc. However, this should be enough fora comparative study. The Instrument functions for the field centre and edge are shown in Fig.~\ref{fig:ifs}.

   These instrument functions are computed for the geometrical width of the slit image on the detector equal to $56 \mu m$ or 4 pixels. It is clear that with this slit the spectral resolution doesn't reach the limits set by the detector and spectrograph's optics. So we consider an alternative case of $40 \mu m$ slit, which may be implemented in practice after the instrument upgrade and installation a variable width slit. Spectral resolwing power, defined by the instrument function's full width at half maximum (FWHM) and the reciprocal linear dispersion  $R= \frac{\lambda}{FWHM dl/d\lambda }$ is presented in Table~\ref{tab:resolution} for both of these cases.
 
\begin{figure} [!ht]
   \begin{center}
   \begin{tabular}{c} 
   \includegraphics[width=0.65\textwidth]{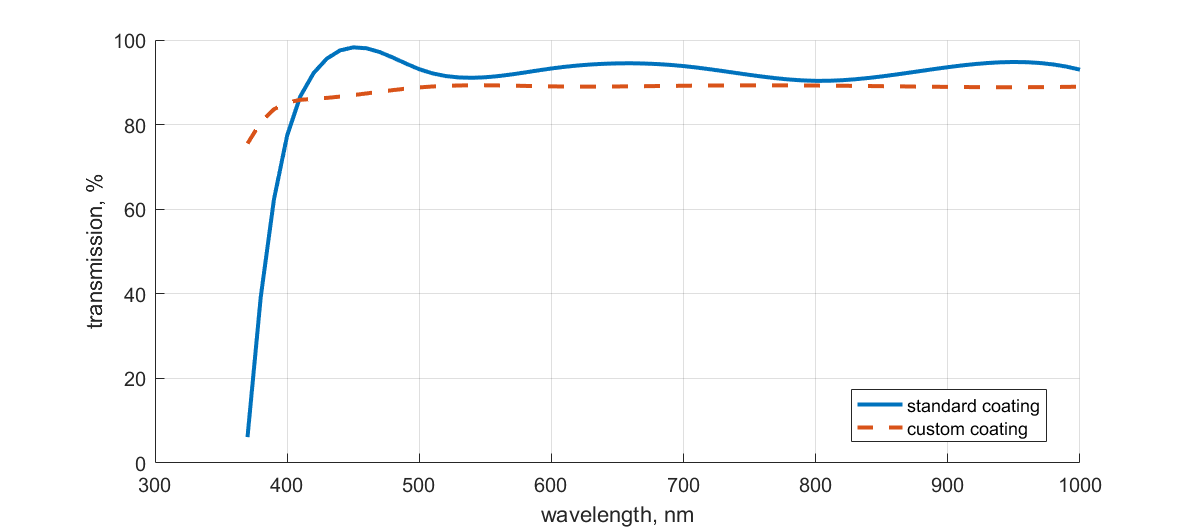}
   \end{tabular}
   \end{center}
   \caption[tau] 
   { \label{fig:tau} 
Spectral dependence of the custom lens transmission with 2 types of AR coating.}
   \end{figure}

   Finally, one of the main reasons behind the development of new camera is its' throughput. With a 5 lens design it becomes possible to reach relatively high transmission even with standard anti-reflection (AR) coatings. For example, a coating equivalent to AB type from Thorlabs\cite{thorlabs}  allows to reach $>90\%$ average transmission for the most of the working range, although it drops sharply at the shortwave edge - see solid curve in Fig.~\ref{fig:tau}. Alternatively, we can consider a customized coating covering the entire waveband. Even a moderate residual reflectivity of $\rho \leq 1\%$ would provide a high and uniform overall transmission - see the dashed curve in Fig.~\ref{fig:tau}.

\subsection{Tolerances estimations}
\label{sec:tol}

In the sake of simplicity we use the geometric aberrations to perform the tolerance analysis. In particular, we set an increase by $10\%$ of the spot diagram RMS radius as the target image quality criterion. Then we perform a standard sensitivity and tolerances analysis for the direct imaging mode, as its' image quality correlates well with that in the spectral modes. The set of variables is typical for a lens design analysis. It should be noted that the aspheres are sensitive to centering and linear decenter is not equivalent to a tilt for them. Finally, only the back focus is used as a compensator. The results are shown in Table~\ref{tab:tol} below in a short form, i.e. only the extreme values per category are specified. Some of these tolerances may be challenging and could require some additional facilities during the manufacturing and alignment, but in general this level of precision is reachable\cite{Langehanenberg}. 

\begin{table}[!ht]
\caption{Minimum and maximum tolerances values for each category. } 
\label{tab:tol}
\begin{center}       
\begin{tabular}{|l|c|c|} 
\hline
Tolerance & Min &Max \\
\hline
Radius , finges @589 nm&	- &	5\\
\hline
Thickness ,$\mu m$	& 15 &	100\\
\hline
Element decenter,$\mu m$  & 15 &	200\\
\hline
Element tilt,$'$  & 0.9 &	11\\
\hline
Aspheres decenter,$\mu m$  & 15 &	28\\
\hline
Surface tilt,$'$  & 0.6 &	6\\
\hline
Surface irregularity,$nm$  & - &	68.3\\
\hline
Refraction index  & $2\times10^{-4}$ &	$1\times 10^{-3}$\\
\hline
Abbe number & 0.147 &	0.95\\
\hline

\end{tabular}
\end{center}
\end{table}

\section{Performance comparison}
\label{sec:compare}  

\begin{figure} [!ht]
   \begin{center}
   \begin{tabular}{c} 
   \includegraphics[width=0.63\textwidth]{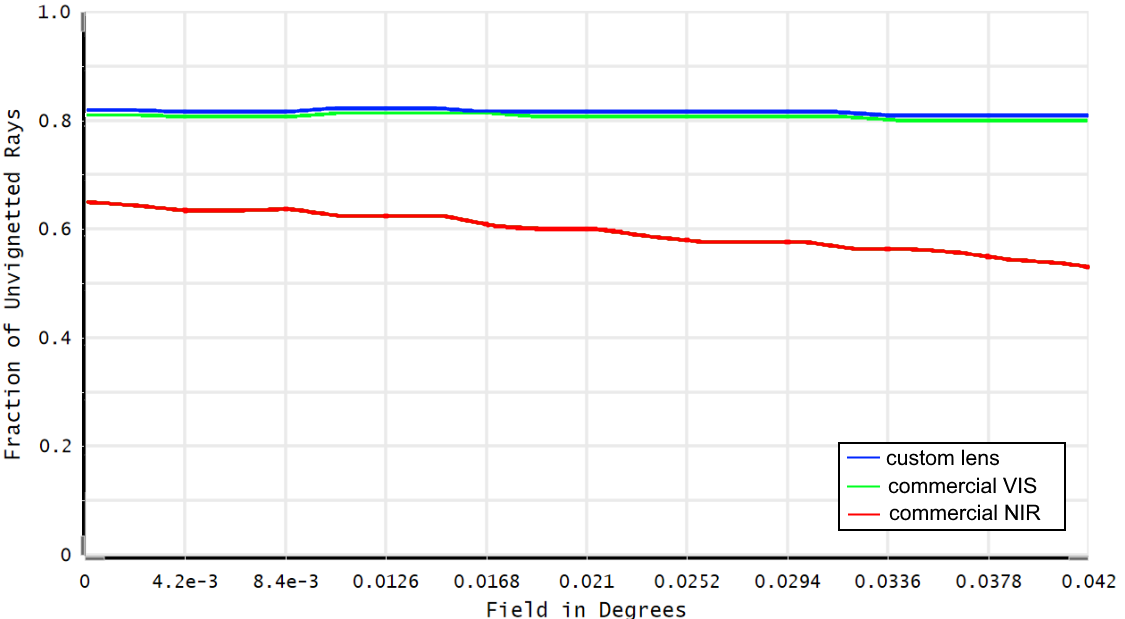}
   \end{tabular}
   \end{center}
   \caption[vig] 
   { \label{fig:vig} 
Comparison of vignetting and central obscuration in the "telescope + spectrograph" optical system with different camera lenses.}
   \end{figure}

First of all, the custom design described above has is superior over the commercial NIR lens as it passes the entire science beam without cropping or field vignetting. Fig.~\ref{fig:vig} visually illustrates the geomteric transmisison gain.

Further, in order to quantify the performance gain reachable with the custom lens, we perform the same computations for the spectrograph equipped with the commercial lenses.

The difference of resolution in direct imaging mode is shown in the summary Table~\ref{tab:direct}. Both the sopt diagram-based metrics and ensquared energy values are shown here. As one can see, in comparison with the Nikkor VIS lens, the proposed design has larger spot wings, but in general the resolution is more uniform over the field. Note, that each time only the nominal waveband is used for simulations, so the chromatism correction for the VIS lens was simpler. At the same time the custom design is superior over the commercial NIR lens regardless of the metric or reference point.

\begin{table}[!ht]
\caption{Resolution metrics for undispersed image} 
\label{tab:direct}
\begin{center}       
\begin{tabular}{|p{2cm}|c|c|c|c|c|c|} 
\hline
Metric &\multicolumn{2}{|c|}{Custom-designed lens}&  \multicolumn{2}{|c|}{Commercial VIS}& \multicolumn{2}{|c|}{Commercial NIR}\\ 
\hline
 & $\omega = 0$ &  $\omega = 2.5'$ &  $\omega = 0$ &  $\omega = 2.5'$ & $\omega = 0$ &  $\omega = 2.5'$ \\
\hline
RMS spot rad.,$\mu m $ & 9.2 & 9.9 & 13.3 &11.1 & 17.9 & 23.0\\
\hline
PTV spot rad,$\mu m $ & 32.6 & 32.6 & 25.6 & 25.1 & 44.8& 75.2 \\
\hline
Esquared energy $r\leq 13.5 \mu m$& $86\%$ & $84\%$& $92\%$& $74\%$& $61\%$& $59\%$\\
\hline
Esquared energy $r\leq 27 \mu m$& $97\%$ & $96\%$& $99\%$& $99\%$& $91\%$& $86\%$\\

\hline

\end{tabular}
\end{center}
\end{table} 

 In a similar way, the spectral resolving power for each camera lens are shown in table~\ref{tab:resolution} to table~\ref{tab:resolution3}. Using the custom lens has a minor effect because of the slit oversampling mentioned before. With the current slit width it provides a modest $\approx 4\%$ growth of the resolution for a part of NIR spectrum.Hovewer, changing the entrance slit may unlock the camera potential and would allow to reach a factor of 1.57 gain in some cases.
 
 \begin{table}[!ht]
\caption{Spectral resolving power with custom lens.} 
\label{tab:resolution}
\begin{center}       
\begin{tabular}{|c|c|c|c|c|} 

\hline
$\lambda, nm$ & \thead{$\omega = 0$,\\ $y'=56 \mu m$} &  \thead{$\omega = 2.5'$,\\ $y'=56 \mu m$} & \thead{$\omega = 0$,\\ $y'=40 \mu m$} &  \thead{$\omega = 2.5'$,\\$ y'=40 \mu m$}  \\
\hline
370	&438&	438&	613	&590\\
\hline
420	&497&	497&	696&	696\\
\hline
500&	592&	592&	829&	768\\
\hline
800&	947&	947&	1326&	896\\
\hline
600&	718&	665&	1005&	1005\\
\hline
800&	947&	957&	1340&	1340\\
\hline
1000&	1184&	1197&	1675&	1675\\
\hline

\end{tabular}
\end{center}
\end{table}

\begin{table}[!ht]
\caption{Spectral resolving power with commercial VIS camera.} 
\label{tab:resolution2}
\begin{center}       
\begin{tabular}{|c|c|c|} 
\hline
$\lambda, nm$ &  \thead{$\omega = 0$,\\ $y'=40 \mu m$} &  \thead{$\omega = 2.5'$,\\$ y'=40 \mu m$} \\
\hline
370	&-&	-\\
\hline
420	&497&	497\\
\hline
500&592&	592\\
\hline
800&947&	947\\
\hline

\end{tabular}
\end{center}
\end{table}

\begin{table}[!ht]
\caption{Spectral resolving power with commercial NIR camera.} 
\label{tab:resolution3}
\begin{center}       
\begin{tabular}{|c|c|c|} 
\hline
$\lambda, nm$ &  \thead{$\omega = 0$,\\ $y'=40 \mu m$} &  \thead{$\omega = 2.5'$,\\$ y'=40 \mu m$} \\
\hline
600&	718&	641\\
\hline
800&	947&	957\\
\hline
1000&	1139&	1197\\

\hline

\end{tabular}
\end{center}
\end{table}

The next metric to compare is the lens transmission. The comparison ignores the geometric vignetting factors discussed before. For the custom lens we use the standard AR coating data (see Fig.~\ref{fig:tau}). For the commercial VIS lens we do not have the exact data, so we have to presume the number of surfaces using the closest known design\cite{Uzawa96} and some residual reflection of the AR $\rho \leq 1\%$. In turn, for the NIR lens the data is known from its' official specifications\cite{Schneider}.  As the table shows, the custom lens is better or comparable in rems of transmission for the most of the waveband and has a significant advantage at the shortest wavelengths. 

\begin{table}
    \centering
    \caption{Lens throughput and resolution comparison}
    \label{tab:Optical_performance}
    \begin{tabular}{|p{2cm}|p{2cm}|p{2cm}|p{2cm}|}
        \hline
         Wavelength, nm &  Custom-designed lens &  Commercial VIS & Commercial NIR   \\
         \hline
         $400$ &  $77.4\%$ &  $18.6\%$ & $52\%$ \\
         \hline
         $500$ &  $93.1\%$ &  $72.5\%$ & $92\%$ \\
         \hline
         $600$ &  $93.3\%$ &  $74.7\%$ & $93\%$ \\
         \hline
         $700$ &  $93.8\%$ &  $75.2\%$ & $93\%$ \\
         \hline
         $800$ &  $90.3\%$ &  n/a      & $95\%$ \\
         \hline
         
    \end{tabular}

\end{table}

Finally, to visualize the advantages of the proposed new camera design, we simulate an image of extended celestial object. We use an image of "Bug Nebula" (Planetary Nebula NGC 6302) for this purpose. It has an angular size of $1.8'$ , which is representative for our nominal field of $\pm 2.5'$. We use the wavelengths $\lambda=420, 600$ and $800 nm$ covering the VIS range instead of the RGB triplet to generate false-colour images. The gain in resolution is clear in fine details of the simulated images. 

\begin{figure} [!ht]
   \begin{center}
   \begin{tabular}{c} 
   \includegraphics[width=0.95\textwidth]{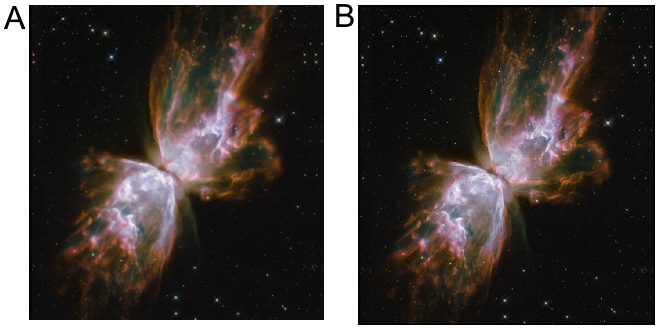}
   \end{tabular}
   \end{center}
   \caption[nebula] 
   { \label{fig:nebula} 
Comparison of simulated images of Bug Nebula: A  -- commercial Nikon lens, B -- custom camera lens.}
   \end{figure} 

\section{Conclusions and future work}
\label{sec:concl}  

In the present paper we proposed a custom camera lens for the MISTRAL spectrograph and imager at 1.93m telescope at OHP. It is shown that this design has significant advantages in terms of throughput and resolution in the direct imaging and long-slit spectroscopy modes. At the same time, the optical design, including the nominal parameters and the tolerances,  is technologically feasible.

The next stages of the project should include development of the opto-mechanical design, which would provide the required accuracy and stability and coupling with the existing interface of Andor camera. The latter point may require some additional effort as the back focal length is shorter than that of the commercial lenses used at the moment. Once the opto-mechanical design is ready, we could proceed to manufacturing, alignment test and installation of the new camera. A simultaneous upgrade of the entrance slit unit would help to maximize the gain in terms of spectral resolving power.

\section*{ACKNOWLEDGMENTS}       
 
Authors thank the CPER OHP2020  Region Sud and the Conseil departemental des Alpes de Haute Provence for their contribution. This research has also made use of the MISTRAL database, operated at CeSAM (LAM), Marseille, France. This work received support from the French government under the France2030 investment plan, as part of the initiative d’Excellence d’Aix- Marseille Universite- A*MIDEX (AMX-19-IET-008). We were also supported by the IPhU Graduate School program at Aix-Marseille University. EJ is a FNRS Senior Research Associate. We also acknowledge the support by Master Erasmus Mundus Europhotonics (599098-EPP-1-2018-1-FR-EPPKA1-JMD-MOB) financed by EACEA (European Education and Culture Executive Agency). Authors thank the CNES for financial support of the MISTRAL operations.

\bibliography{main}
\bibliographystyle{ieeetr}

\end{document}